\documentclass[english]{aa}
\usepackage{mathptmx}
\usepackage[T1]{fontenc}
\usepackage[latin9]{inputenc}
\usepackage{float}
\usepackage{textcomp}
\usepackage{amstext}
\usepackage{graphicx}
\usepackage{amssymb}

\makeatletter

\providecommand{\LyX}{L\kern-.1667em\lower.25em\hbox{Y}\kern-.125emX\@}
\DeclareRobustCommand*{\lyxarrow}{%
\@ifstar
{\leavevmode\,$\triangleleft$\,\allowbreak}
{\leavevmode\,$\triangleright$\,\allowbreak}}
\newcommand{\lyxmathsym}[1]{\ifmmode\begingroup\def\b@ld{bold}
  \text{\ifx\math@version\b@ld\bfseries\fi#1}\endgroup\else#1\fi}

\providecommand{\tabularnewline}{\\}

\makeatother

\usepackage{babel}

\begin{document}

\title{Probing dark matter haloes of spiral galaxies at poorly explored
  distances using satellite kinematics\thanks{Based on observations
    made with ESO Telescopes at the la Silla and Paranal Observatories
    under programmes 075.B-0794 and 077.B-0767.}}


\author{I.A. Yegorova\inst{1}\and A.Pizzella\inst{2}\and P.Salucci\inst{3}}

\institute{European Southern Observatory, Alonso de Cordova 3107, Santiago,
Chile \email{iyegorov@eso.org}\and Dipartimento di Astronomia, Universit\`{a}
di Padova, Padova, Italy \email{alessandro.pizzella@unipd.it} \and SISSA
International School for Advanced Studies, via Beirut 4, I-34013 Trieste,
Italy \email{salucci@sissa.it}\\
}

\offprints{I.A. Yegorova}

\date{Received ...; Accepted...}

\date{}

\abstract{}{We present the results of a pilot project designed to
  study the distribution of dark matter haloes out to very large
  radii in spiral galaxies. As dynamical probe we use their rotation
  curves and the motions population of satellite galaxies. In this pilot 
  stage, we observed seven late-type spiral galaxies of about the same luminosity
  $M_R \sim -22$ (and approximately the same mass). We investigate the
  kinematics of these galaxies, and the radial and angular distribution 
  of their satellites.}  {Using VIMOS at the VLT, we carried out a spectroscopic
  survey in seven $14'\times 14'$ fields each around a late-type isolated
  spiral galaxy. We obtained radial velocities and spatial
  distributions for 77 candidate satellites. After removing the interlopers, 
  we are left with  61 true satellites. In combination with the rotation curves of the
  primary galaxies, satellites are used to probe the gravitational
  potential of the primaries and derive the dark matter halo properties by means 
  of standard mass modeling techniques.}  {We find (a) that the dark
  matter haloes of luminous spirals ($M_R \sim -22$) 
  have virial radii of $\sim 400$ kpc and virial 
  masses of $3.5 \times 10^{12} \rm  M_{\odot}$; (b) 
  that the radial velocity and angular distributions of
  the satellites around the primaries are isotropic; and (c) that the
  resulting mass distribution is in good agreement with that found in
  the optical regions of spirals and described by the universal
  rotation curve of spirals once extrapolated to large radii. The
  results obtained in this pilot phase of the project are already interesting
  and limited only by small number statistics. The full project
  involving an order of magnitude more targets, would
  very likely provide us with a definitive picture of the dark matter
  distribution around spirals out to their virial radii and beyond.}  {}


\titlerunning{Probing dark matter haloes of spiral galaxies at poorly explored
  distances using satellite kinematics}
\maketitle


\section{Introduction}

In the past few decades, it has been revealed that dark matter (DM) haloes
extend far beyond the optical boundaries of galaxies. However, their
actual ``sizes'' and, above all, their density distribution outside
the luminous regions, are still quite unknown. The inner regions of DM
haloes are studied using several methods including the analysis of both
rotation curves for spiral galaxies (Rubin et al. \cite{Rubin}; Bosma
\cite{Bosma}; Persic, Salucci \& Stel \cite{Persic&Salucci}), and
velocity dispersions for ellipticals (Kronawitter et al.
\cite{Kronawitter}; Cappellari et al. \cite{Cappellari}; Thomas et al.
\cite{Thomas}). Different tracers of the gravitational field,
must be used to investigate the external regions, which are relatively void of
luminous matter. These include planetary nebulae (Coccato et al.
\cite{Coccato}) and globular clusters in ellipticals (Spitler et al.
\cite{Spitler}), weakly lensed background galaxies, and the satellites
populating the outer parts of the DM haloes. These objects are likely to be
the remnants of the halo formation process and provide us with an
efficient dynamical probe, out to the ``edge'' of the halo and beyond,
as they are bright, numerous, and radially extended test particles.  
By measuring the satellite--primary relative velocity 
for a sufficient number of objects, we can
derive the mass distribution of the halo out to very large radii.
The results are complementary to those obtained from the
analysis of weak lensing data, not only because the measurements involved
and their related biases are independent, but 
because the satellite motions lead to a measure of  $V(R)= (GM_h(R)/R)^{0.5}$ 
i.e. the halo mass inside a given radius, while the observed tangential shear
derived the weak lensing measures a more complex quantity, the (average
- local) halo surface density at a given radius $\bar \Sigma(<R)$ -
$\Sigma(R)$.  Since $(GM_h(R)/R)^{0.5}$ in the regions under
study at most decreases as $R^{-0.25}$, the sensitivity of the
dynamical measurements performed using satellite kinematics does not
depend on radius, in contrast to what occurs for the majority of tracers
of the gravitational potential including weak lensing. Finally,
comparing different mass models derived from various kind of measurements, 
provides a valuable consistency check.

The virial radius $R_{vir}$ is an important characteristic scale-length of dark matter haloes. 
In the current concordance cosmology (where $\Omega_{CDM} \sim  0.3$, $\Omega_{\Lambda} \sim 0.7$,  
and $h= 0.72$ represent the dark matter, dark energy densities, and Hubble parameter, respectively),
is defined as the radius that encompasses 330 times the background mass  
$ 4/3  \pi R_{vir} ^3 \Omega_\mathrm{CDM} \rho_c$, with $\rho_c \sim 10 ^{-29} \ g/cm^3 $ 
the critical density of the Universe. For historical reasons $R_{200}$, 
i.e. the radius that encompasses a mass  $ 4/3 \pi R_{vir} ^3 200 \rho_c$, is also often considered, 
where is related to $R_{vir}$ all halo structural properties (concentration and mass) and can be taken 
as the reference size of a DM halo (e.g. Navarro et al. \cite{Navarro}). Moreover, simulations show  
that it approximately marks the transition between the virialized matter of a halo and the
infalling material. In detail, for a virialized structure of mass $M_{vir} $, at redshift zero, 
$R_{vir} = 260(M_{vir}/(10^{12} M_{\odot } ))^{1/3}$ kpc defines the typical scale-length 
relevant to this study.

From an observational point of view, the distribution of the DM
outside the inner luminous regions out to $R_{vir}$ is still poorly
known, and data suitable for investigating these issues are scarce. 
Invaluable data is provided by weak lensing measurements (e.g. Mandelbaum
\cite{Mandelbaum}), which indicate that $R_{vir} $ is some hundreds of
kpc, although the precise value is uncertain because of the weakness of the signal 
at these distances coupled with the uncertainties in the DM density profile.

In this pilot study, we show that it is possible to investigate
the DM distribution of spiral galaxies out to $R_{vir} $ by combining the internal
kinematics of the spirals and the radial velocities of their satellites.

It is well known that the spatial and kinematical distribution of satellites 
around their host galaxies provide important information about the DM halo 
properties. Previous investigations have also noted that
further analysis is possible only if the angular distribution of satellites 
is homogeneous. It is extremely difficult to recover the  gravitational 
potential of the galaxy halo from  the  kinematics of satellite systems if their 
distribution is not (roughly) isotropic. Zaritsky et al. (\cite{zaritsky_97}) studied 
a sample of $69$ spiral galaxies  and their $115$ satellites, and found 
a tendency for the satellites to lie preferentially close to the minor
axis of their hosts, a phenomenon called the Holmberg effect (Holmberg
\cite{Holmberg}).  Brainerd (\cite{brainerd_05}; hereafter B05)
studied a sample of 2000 primaries and their 3300 satellites selected
from the Sloan Digital Sky Survey Data Release 3 (SDSS DR3; York et al.
\cite{York}), and found instead a tendency for satellites to lie
near the major axis, a result confirmed by Faltenbacher et al.
(\cite{Faltenbacher}) using SDSS DR4 data. Agustsson \& Brainerd (\cite{Agustsson}) 
studied a large sample of SDSS DR7 galaxies (4487 hosts and 7399 satellites) and 
for blue hosts found that at projected radii $r_{p} \lesssim 150$~kpc
satellites are distributed close to the major axes, while at $r_{p}
\gtrsim 300$~kpc satellites are found close to the minor axes. Azzaro
et al. (\cite{Azzaro}) found an isotropic distribution of satellites
based on a sample of 193 satellites and 144 isolated spiral galaxies,
though they issued a caution about the insufficient quality of their statistics. 
Using SDSS DR5 data, Bailin et al. (\cite{Bailin}) confirmed this result with
$273$ late-type primaries hosting $321$ satellites.

We believe that the contradictory results found so far have the
following causes. First, most of these studies rely on data for a very small
number of satellites per host. The average primary galaxy often
possesses data for only two satellites, which is a very small fraction of the
true total number, as we can see just by looking at the Local Group
(Hartwick \cite{Hartwick}; Kroupa et al. \cite{Kroupa}; Willman et al.
\cite{Willman}; Koch et al. \cite{Koch}). Second, to combine  the
kinematics of systems composed of primaries of different masses and
halo sizes is very complex and the subsequent analysis is far from
trivial and can become easily biased.

This uncertain situation has suggested to us a new observational approach 
to the problem. First, we select a relatively small number of primary 
galaxies (compared to previous studies), and all in a quite
narrow range of luminosity $-21.6 \leq M_R \leq -22.7$. For each
of them, we then detect a larger number of satellites by using powerful 
instruments (VIMOS at the VLT). The characteristic halo mass will 
eventually be sampled by 30 primaries in that luminosity range, 
and their gravitational potential will be reconstructed by means of a)
the disk kinematics and b) the kinematics of their about 300  possibly detected
satellites.

In this pilot paper, which is also a feasibility study, we carry out about 25\% of 
the project. To complete the project and investigate the mass distribution around
spirals of different masses, we plan to replicate the investigation around objects 
of significantly lower luminosity (e.g. with $-20 \leq M_R \leq -20.5$).

We measure the rotation curves (RCs) of seven primaries to help us reconstruct the 
mass distribution in the inner regions of galaxies and eventually ensure 
that the systems have approximately the 
same total mass ($2-5 \times 10^{12}~\rm M_\odot$). 
We then measure the primary-satellite relative velocities
for the candidate satellites and, after removing interlopers,  we build
for the latter the velocity dispersion profile of the co-added system.
The method has a number of advantages: a) Since the primaries have
almost the same mass, we can co-add the kinematical measurements from
different objects as they would belong to a single one. This leads to
a very precise measure of the rotation curve in the optical region, a
good hint of the satellite number profile, and a fair measure of their
velocity dispersions. b) Since we  detect objects down to a very low 
($M_R = -16$) limiting magnitude, our dynamical analysis is not affected
by possible luminosity-dependent biases in the positions and in the
motions of satellites.  c) We can easily reach
the virial radius and beyond.

In this paper, we report the first results of the pilot phase
of our project, obtained with seven hours of VIMOS at the VLT, followed
later by a program totaling 53 hours at the VLT (VIMOS) and NTT
(EMMI). In this phase, we obtain valuable information about the properties at the outer
radii of dark and luminous components around spirals, but we are also
particularly interested in demonstrating the feasibility of the project. We
anticipate that, once completed, our project will be able to unravel most 
of the distribution of DM within the haloes surrounding spiral galaxies out to
their ``edges''.

The layout of the paper is the following. In Section 2, we describe
observations and data reduction of primary galaxies and their
satellite candidates. In Section 3, we perform the data analysis: in
Section 3.1 of the structural properties of the primaries, using their
rotation curves; in Section 3.2 identifying interlopers in the
satellite candidates; in Section 3.3 studying the radial distribution
of satellites; in Section 3.4 studying the angular distribution of
the satellites; in Section 3.5 estimating the velocity dispersion
profile; and in Section 3.6 performing mass modeling of the primary
galaxies. We discuss the results in Section 4.

\section{Observations and data reduction}

\begin{table*}
\begin{center}
\caption{Primary galaxies characteristics and observing blocks. 
\label{tab:Primary galaxies}}

\noindent 
\begin{tabular}{ccccc}
 &  &  &  & \tabularnewline
\hline
\hline 
Primary galaxy  & z  & mag$^{\rm a}$ & Exp.Time EMMI$^{\rm b}$ & Exp.Time VIMOS$^{\rm b}$\\
\hline 
J003828+000720      & 0.042 & 14.60 & 2 x 2200       & 2 x 2180\\
J134215+015126      & 0.029 & 13.60 & 2 x 2000       & 5 x 2180\\
J145211+044053      & 0.046 & 14.62 & 3 x 2000       & 6 x 2180\\
J152621+035002      & 0.083 & 14.94 & 2 x 2000       & 5 x 2180\\
J153221-002549      & 0.085 & 15.18 & 2 x 2200       & 8 x 2180\\
J154040.5-000933.5  & 0.076 & 15.70 & 4 x 2700 +1800 & 6 x 1000\\
J154904-004023      & 0.077 & 15.73 & 2 x 2200       & 5 x 2180\\
J221957-073958      & 0.038 & 13.57 &           1800 & 5 x 2180\\
\hline
\end{tabular}

\smallskip
$^{\rm a}${\scriptsize $r$-band magnitudes from SDSS}

$^{\rm b}$ {\scriptsize exposure time is in seconds}
\end{center}
\end{table*}

\smallskip{}

The selection criteria of the primary galaxies were the following:
(1) all galaxies are late-type spirals; (2) they have an inclination
between 40\textdegree{} and 70\textdegree{}; (3) they are isolated:
all objects within a radius of $700$ kpc and $1000~{\rm km~s^{-1}}$ 
in relative radial velocity have a luminosity at least 2.5 magnitudes 
below that of the primary galaxy; (4) they have a redshift $0.029<z<0.085$; and 
(5) they have at least four known satellites (from the SDSS). To select the isolated
primaries, we used the catalog of host galaxies and satellites from B05.

The observations were conducted with two instruments: the kinematics
of primary galaxies were measured using data acquired with EMMI at the NTT, while the
hunt for satellites and the measurement of their radial velocity was
done with the VIMOS at the VLT. Redshifts and luminosities of the primary
galaxies, and the observing log, are presented in Table~\ref{tab:Primary galaxies}.

\subsection{Kinematics of primary galaxies}

\begin{figure}
\begin{raggedright}
\includegraphics[width=1.0\columnwidth]{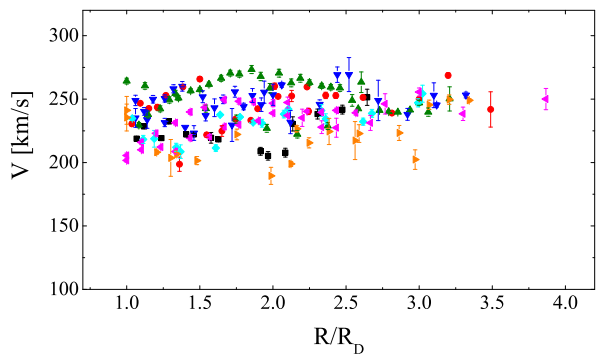}
\par\end{raggedright}

\caption{Rotation curves of the seven primary galaxies deprojected for the
 disk inclination and folded around their centers.
\label{fig:RC}}

\end{figure}

The major-axis ionized gas rotation curves of the primary galaxies
were obtained by means of long-slit spectroscopy at the ESO-NTT~3.5m
telescope in La Silla (077.B-0767), using the ESO Multi-Mode Instrument
(EMMI) in the REd Medium Dispersion (REMD) configuration. Grating
No.~6 with 1200~grooves~mm$^{-1}$ blazed at 6890~\AA\ was used
in the first order in combination with the two red-arm mosaicked MIT/LL
and a 1\arcsec$\times$330\arcsec slit. Data reduction was
done with MIDAS\footnote{MIDAS is developed and maintained by ESO.} 
and IRAF\footnote{IRAF is distributed by the National Optical Astronomy
Observatories which are operated by the Association of Universities for 
Research in Astronomy under cooperative agreement with the National Science
Foundation.} standard routines. Wavelength calibration was performed using
observations of a helium-argon calibration lamp. All frames were bias subtracted,
flat-field corrected, cosmic-ray cleaned, and wavelength calibrated.
Quartz lamp and twilight sky flat-fields were used to remove pixel-to-pixel
variations and large-scale illumination patterns. After wavelength
calibration, repeated exposures of the same target were aligned
and summed. We used the night-sky emission lines to
assess the quality of the wavelength calibration and spectral resolution.
We use these data not only to derive the rotation curve
of the galaxies but also to define their recession velocity. This
velocity is later compared with the one measured for the satellites
using VIMOS. The instrumental resolution is 1.3~\AA\ (FWHM), corresponding
to a resolution of $\sigma=27~ {\rm km~s^{-1}}$ at H$\alpha$.
The ionized gas kinematics was measured by simultaneous Gaussian fits
to H$\alpha$, {[}\ion{N}{ii}{]}, and {[}\ion{S}{ii}{]}\ lines by
means of a self-written IDL routine and then folded around their centers
and deprojected for the disk inclination as done by Pizzella et al. 
(\cite{Pizzella}).  The derived velocity curves are plotted in 
Fig.~\ref{fig:RC}.

\subsection{Satellites radial velocity}

Observations were carried out in service mode with the VIsual MultiObject
Spectrograph (VIMOS) operated at Cerro Paranal (075.B-0794; 077.B-0767).
To identify satellites orbiting selected primary galaxies and measure their
radial velocity, the instrument was used in MOS configuration. After
pre-imaging, we constructed the masks for the MOS observations selecting
the brightest non-stellar objects in the field.

We adopted grism ``HR orange'' on the four arms of the instrument.
The grism is used in combination with the order sorting filter GG435.
The grism is characterized by a dispersion of $0.6$~\AA~px$^{-1}$, which
corresponds to a spectral resolution of 2150 for a 1'' slit. We could 
accommodate, on average, about 50 slits in each quadrant for a total of
about 200 slits for each primary galaxy. We organized the observations
by repeating the same pointing with the same mask several times. Each
exposure is 2180 seconds long. Some exposures that did not fulfill
our observational requirements were repeated. For this reason,
we obtained more exposures than requested. Extra exposures, executed
under weather conditions not complying with our constraints, were
only used if the data quality was good. The exposures reported
in Table~\ref{tab:Primary galaxies} are the ones used in our final
data reduction.

We started our data reduction from frames that were bias subtracted,
flat-field corrected, and wavelength calibrated by the ESO data reduction
pipeline. Further data reduction was done with MIDAS. The first
step was to assess the quality of each integration. Once the selection
was done and the low quality spectra discarded, we computed the average
spectrum for all frames. The second step was to approximately derive,
in the sky-subtracted spectra, the redshift of all targets. To this
aim, we used emission and absorption lines listed in Table~\ref{tab:lines}.

\begin{table}
\caption{Emission and absorption lines used in the preliminary redshift 
determination.\label{tab:lines}}

\noindent \centering{}\begin{tabular}{cc}
\hline
\hline 
Line           & Wavelength (\AA ) \\
\hline 
{[}\ion{O}{ii}{]} & 3726.03\\
{[}\ion{O}{ii}{]} & 3728.82\\
\ion{Ca}{ii} K    & 3933.66\\
\ion{Ca}{ii} H    & 3968.47\\
H$\delta$         & 4101.73\\
H$\gamma$         & 4340.36\\
H$\beta$          & 4861.31\\
{[}\ion{O}{iii}{]}& 4958.92\\
{[}\ion{O}{iii}{]}& 5006.84\\
Mg                & 5183.6 \\
H$\alpha$         & 6562.80\\
{[}\ion{N}{ii}{]} & 6583.41\\
{[}\ion{S}{ii}{]} & 6716.47\\
{[}\ion{S}{ii}{]} & 6730.85\\
\hline
\end{tabular}
\end{table}

This second step allowed us to select possible satellites and reject
all other background or foreground objects characterized by 
a radial velocity difference  (with respect to the primary galaxy) 
more than $V_{out} = 1000~{\rm km~s^{-1}}$ (McKay et al. \cite{McKay}; 
B05). We refer to these galaxies as {\it candidate satellites} 
throughout the paper.

In designing the masks containing the slits, we did not
constrain the wavelength range. Slit
positioned on the north side of each quadrant produced spectra in
the red range, with wavelengths longer than 5500~\AA . Slit positioned
on the south side of each quadrant produced spectra in the blue range,
with wavelengths shorter than 6500~\AA . Our strategy is to derive
the redshift of satellites either from the H$\alpha$, {[}\ion{N}{ii}{]}\ 
emission lines or from the H$\beta$, {[}\ion{O}{iii}{]}\ emission lines
that always fall in the observed spectral range. We were typically able to
identify about 150 redshifts for each primary galaxy. 

After the satellite candidates of a primary galaxy were identified,
we proceeded with the accurate measurement of their radial velocity. We again,
used all available emission lines (typically H$\alpha$, {[}\ion{N}{ii}{]}\ 
or H$\beta$, {[}\ion{O}{ii}{]}) and used the sky emission lines to
refine the wavelength calibration. We first spatially averaged the
spectrum of each satellite and derived its one-dimensional spectrum.
We then measured the wavelengths of a number of night-sky and target
emission lines. We later, computed the difference between the expected
night-sky wavelength values and the measured ones, which were fit
with a one-degree polynomial. This was used to correct the wavelength
of the satellite emission lines. To have independent measurements,
when possible we measured the radial velocity separately for each
exposure and for each emission line and then averaged the measurements.
The radial velocities were measured with a typical uncertainty of
$10~{\rm km~s^{-1}}$. All radial velocities were reported with respect 
to the heliocentric rest-frame.

The last step consisted of identifying the positions of all the 200
slits on both the pre-imaging frame and the SDSS images and, in particular,
the positions of the satellites.

In view of future observations with the same instrument, we note, 
that we can easily increase the fraction of observed
extragalactic sources, and therefore satellites, by using different
MOS masks for the same primary instead of repeating the same mask
several times as we did in this pilot project. In very few cases, 
we added the spectra before measuring the redshift and we found that 
two independent $45$ min spectra per target would have provided a velocity 
measurements with uncertainties not significatively larger than what we 
obtained with several exposures. 

\section{Distribution of satellites}
\begin{figure}[t]
\begin{raggedright}
\includegraphics[width=0.8\columnwidth]{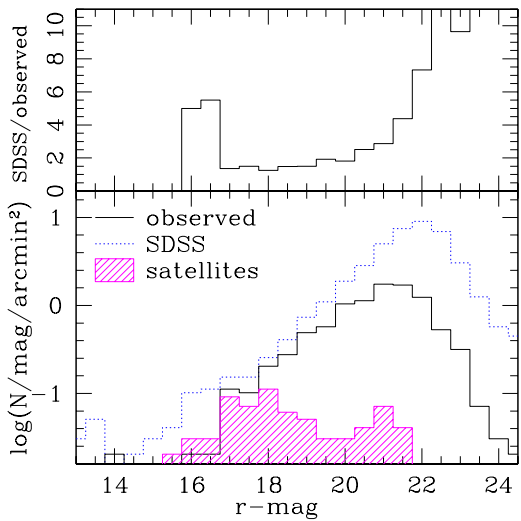} \par\end{raggedright}

\caption{Lower panel: SDSS DR7 number counts of galaxies averaged in a
  $14'\times 14'$ field around each primary galaxy ({\em blue dotted
    line}), number counts of the galaxies that have been observed with
    VIMOS ({\em black line}, magnitudes from SDSS), and number counts of
    the physical satellites ({\em shaded region}). Upper panel: ratio 
    of the number of galaxies present in the field to the one observed 
    with VIMOS. 
\label{fig:completness}}

\end{figure}

At the end of the data reduction phase, we found a total of 77 objects
(considering both VIMOS and B05 data) that fall in a projected circle
in the sky of about 1000 kpc of radius and have a velocity that is at most
$ 1000~{\rm km~s^{-1}}$ smaller or larger than that of the primary galaxy. 
In Table~\ref{tab:number}, we indicate the number of  satellite candidates found 
for each primary galaxy (for J003828+000720 the quality of the data was  
insufficient for the analysis). Not all objects are physical satellites
since outliers may be present (see Sec.\ref{subsec:interlopers}). 
To assess the completeness of our sample, we estimate the fraction of objects 
that we observed over the total number of objects present in the field.
For each VIMOS field, we derived the $r$-band magnitude
distribution of all extended sources, regardless of their redshift,
from the SDSS DR7 photometric catalog. We then identified in the latter 
the objects that we had observed and found that, in the magnitude range 
between 18 and 21 $r$-band, they are about half of those present in the SDSS 
photometric catalog (see Fig.~\ref{fig:completness}) and this ratio is roughly 
constant with the apparent magnitude. Only for objects fainter than 21 the ratio start 
to increase. The discontinuity visible at magnitude $\sim 16-17$ is caused by our having 
for these objects an SDSS redshift and we selected only the few galaxies with 
a recession velocity similar to the one of the primary galaxy. We can assume that,
by adopting different masks in our observations, we could
spectroscopically observe twice the number of objects we actually
observed.  Since we placed the slits on targets with no particular
criteria other than that they are extended sources, the real number of candidate
satellites (galaxies with a radial velocity $V_{\rm r}$ in the interval
$V_{primary}-1000~{\rm km~s^{-1}} < V_{\rm r} < V_{primary}+1000~{\rm km~s^{-1}}$)
present in the field is about twice the number we counted. 
We have to take into account this factor of about two when estimating
the number density of objects.

\begin{table}[h]
\caption{Number of candidate and confirmed satellites found for each galaxy. \label{tab:number}}

\noindent \begin{raggedright}
\begin{tabular}{cccc}
&  &  & \\
\hline
\hline
ID & SDSS$^{\rm a}$ & SDSS+VIMOS$^{\rm b}$ & Satellites$^{\rm c}$\\
\hline
J134215+015126   & 4 & 8  & 7\\
J145211+044053   & 4 & 7  & 7\\
J152621+035002   & 4 & 17 & 16\\
J153221-002549   & 5 & 19 & 5\\
J154040.5-000933 & 7 & 11 & 11\\
J154904-004023   & 2 & 8  & 8\\
J221957-073958   & 4 & 7  & 7\\
\hline
\end{tabular}
\par\end{raggedright}

\smallskip

$^{\rm a}$ {\scriptsize number of candidate satellites known from SDSS (BO5).}

$^{\rm b}$ {\scriptsize candidate satellites from SDSS plus VIMOS observations.}

$^{\rm c}${\scriptsize satellites after rejection of interlopers.}
\end{table}

\subsection{Structural properties of the primaries from their rotation curves}
\label{subsection:RCs}

{A crucial advantage of this study is that by measuring the RCs of the primaries, 
we can estimate their inner gravitational field. As a first step, from their velocity--position diagrams 
and inclinations $i$, we derive their corrected rotation curves. Secondly, we compute the average 
velocity of all curves at radius $R_\mathrm{opt}=3.2\,R_\mathrm{D}$ 
(where $R_\mathrm{D}$ is the exponential disk scale-length). We define this velocity $\left\langle V(R_\mathrm{opt})\right\rangle$, 
and we use it to normalize every RC to a same representative amplitude and radius. We found 
$\left\langle V(R_\mathrm{opt})\right\rangle = (255 \pm 16)~{\rm km~s^{-1}}$ and $\left\langle R_\mathrm{opt}\right\rangle = (19.8\pm 2.0) $~kpc. 
Because of the small luminosity range of the primaries, both quantities lie in a relatively
small range of values,  in agreement with the Tully-Fisher relation. 
Following  Persic, Salucci \& Stel (\cite{Persic&Salucci}; hereafter PSS), for each primary we plot
$V(R \left\langle R_\mathrm{opt}\right\rangle/R_\mathrm{opt})~  \left\langle V(R_\mathrm{opt})\right\rangle/V(R_\mathrm{opt})$ 
(see Figs.~\ref{fig:bin-sat1} and ~\ref{fig:bin-sat2}), a quantity which 
is proper when  comparing rotation curves of objects of (slightly) different luminosity.  
We find, in agreement with PSS and Salucci et al. (\cite{URCII}; hereafter S07), that  
the optical  RCs of the  primaries  have similar shapes and define universal  
curve $V_{URC}(R/R_\mathrm{opt}, L)$ that accurately represents each of them. This implies that 
these galaxies have similar disks embedded  in similar dark matter haloes. 
Therefore their RCs and the recessional velocities of their satellites 
probe the dark matter halo  around a  spiral of luminosity $\left\langle M_R\right\rangle$
with characteristic radius $\left\langle R_\mathrm{opt}\right\rangle$. 
The virial mass, virial radius, and halo escape velocity can roughly be estimated 
by extrapolating its representative RC out to large radii according to $V(R)=V(R_\mathrm{opt})(R/R_\mathrm{opt})^{-0.2}$, 
i.e.  by adopting the velocity profile suggested by 
observations in S07 and found in $\Lambda$CDM simulated haloes.

We find $M_{vir}\sim 3.5 \times 10^{12} \rm M_{\odot}$ and $R_{vir}\sim 400~$kpc, which we consider as 
indicative values of the virial masses and radii of the primaries. Furthermore,
we can estimate the escape velocity as $V_e(r)=\int_r^\infty V(r)^2 / r \ dr$, 
because it mainly depends on the gravitational potential at small
radii, which can be accurately mapped by the (co-added) RC. This estimate is robust in
spite of the uncertainty in the RC extrapolation: even if we truncated
the halo at 1/2 $R_{vir}$, $V_e(r)$ would be smaller by only 20\%.
We also note that the value of $V_e(r)$ varies with halo mass only as $M_{vir}^{1/3}$.

We use these estimates just as templates of these quantities and $V_e(r)$ helps us to 
distinguish interlopers from true satellites. We note that the actual  circular velocity curve 
leading to the gravitational potential  for our galaxies will be derived from the velocity 
dispersion of the satellite population. No result of this paper depends on the details 
of the above RC extrapolation.   

\begin{figure}[t]
\begin{raggedright}
\includegraphics[width=0.8\columnwidth]{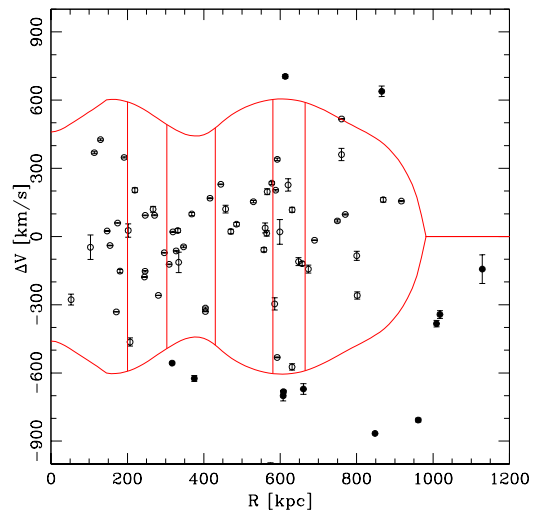} \par\end{raggedright}

\caption{Satellites velocities with respect to the primary ({\em open circles}) versus radius. 
{\em Black points} are outliers. The {\em red curves} are the caustics identifying satellites and outliers. {\em Vertical} 
red lines indicate the bin locations and sizes.
\label{fig:dv}}

\end{figure}

 
\subsection{Interlopers}
\label{subsec:interlopers}

Not all the galaxies detected around the seven primaries are their
satellites: the distribution of objects around a target galaxy is
contaminated by the presence of a background population of
interlopers/outliers, i.e. of objects physically unrelated to the
galaxy targets but at a small projected distance from them. 
That we investigate haloes of approximately the same mass and virial
radius, and that these quantities are roughly known, allows us to infer
the membership status of nearby objects according to their location
and relative velocity with respect to the primary. Moreover, we can also 
distinguish the background of "false satellites" in outliers, galaxies 
around the primaries whose large value of relative velocity $V_{rel}$ show 
that they are systems unbound to the primaries, and true velocity
interlopers, i.e. objects with $V_{rel}$ and projected locations
similar to those of true satellites.

In general, the removal of interlopers is quite complicated and
uncertain. However, according to Chen et al. (\cite{chen}) and Azzaro
et al. (\cite{Azzaro}) for haloes of $10^{12}$ - $10^{13}~\rm
M_{\sun}$, the removal of outliers can be routinely done and the
contamination of velocity interlopers becomes important only at
projected radius of $R> 400$~kpc $\sim R_{vir}$. 

A modern and efficient way of identifying member satellites is the caustic method
introduced by Diaferio (\cite{Diaferio}) (later improved in Serra et al. (\cite{Serra10}; 
\cite{Serra11})). The method uses relative velocities $V_{rel}$ and projected radii to 
delineate the caustic curves that represent the escape velocity of the system (for 
details, see Serra et al. (\cite{Serra10}; \cite{Serra11})).
We applied it to our data identifying the interloper galaxies and removed 
them from our sample. In Fig.~\ref{fig:dv}, we present the $V_{rel}$ versus $R$ for all objects, 
along with  the caustic defined by  Diaferio (\cite{Diaferio}).
We also show the locations of the bins adopted (ten objects per bin) in computing the satellites
mean velocity and velocity dispersion.

Since the present work is limited by the relatively small number
of satellites, we investigate this  issue  by means of a further  method  to identify objects 
gravitationally unbound to our primaries. We also used the negative binding energy criterium: 
we resorted to the negative binding energy criterium that uses the $V_{e}(r)$ profiles by obtaining 
a very similar result in the selection.

With the above considerations, we removed from our sample, at any
projected radius $R$, the unbound/unrelated objects to the primaries
whose inclusion strongly affects the estimate of the velocity dispersion.
These outliers  are  galaxies  more distant  than $5$ Mpc from the primaries,
which happen to lie  on a line of sight passing close to their centers (i.e. $<1000$~kpc).
In Fig.~\ref{fig:densita2}, we give the histogram of our objects after we performed the velocity 
clipping described above, i.e. we show  the number of candidate satellites for each radial 
bin and the number of outliers. Because of small numbers, the cosmic variance of outliers 
cannot be estimated in this study.

The clipping  does not remove objects  that are physically 
unrelated to the  primaries, but reside at a smaller distance
($1$ Mpc $<D<5$ Mpc). In this case, $V_{rel}$,  
which is the resultant of redshift and peculiar velocity,
can take similar values in a true satellite or in a physically
unrelated object. This population of velocity interlopers affects
in particular the satellite's surface density (at outer radii); we
investigate its bias of our measured projected satellite surface
density by following Azzaro et al. (\cite{Azzaro}) and assuming
that it provides a constant background (Chen et al. \cite{chen},
Azzaro et al. \cite{Azzaro}, Mamon et al. \cite{mamon}). We estimate
the latter with the SDSS $r$-band luminosity function (Blanton et al.
\cite{blanton}), i.e.  a Schechter function $ \Phi(M) = 0.4\Phi_{*}
\log(10)\ 10^{0.4(M_{*} - M)(\alpha + 1)}\exp({-10^{0.4(M_{*} -
    M)}})$, with the parameters $\Phi_{*} = 0.0074$ Mpc$^{-3}$, $M_{*}
= -21.31$, and $\alpha = -0.89$.

In detail, we estimate the number of velocity interlopers $N_{vi}(R)$, i.e.  
of the objects that lie within a projected distance $R$ from the center of 
a primary galaxy but at a redshift distance $D_{e} \leq V_{e}/72$ Mpc.
Assuming that $V_e$ is constant with radius, we get 

\begin{equation}
\begin{array}{cc}
N_{vi} = \int_{-14.5}^{-20} \Phi(M) dM\ \int^R_0 2 \pi  V_{e}(R'/72) R' dR' \vspace*{0.2cm} \\
 = 1  \times (R/400~{\rm kpc})^{2},\end{array}
\label{eq:Nvi}
\end{equation}
where $M_R=-14.5$ is the limiting magnitude in our observations, and $72~{\rm km~s^{-1}}$ Mpc$^{-1}$ is the Hubble constant. 
In agreement with Azzaro et al. (\cite{Azzaro}), we consider the radial dependence of $V_e$ (see Sec. \ref{subsection:RCs}) and find that
$N_{vi} \sim 1 \times (R/1$ Mpc$)^{1.75}$ i.e.  in our case the number of interlopers inside $R_{vir} \simeq 400$~kpc is at most one, 
while in the region between $R_{vir}$ and $2 R_{vir}$ (that corresponds to the three outermost radial bins) the expected number 
of interlopers is a few. In terms of number surface density, we find that $\mu_{vi} \sim 5 \times 10^{-7}$ kpc$^{-2} $, with an uncertainty 
of a factor of two.  Thus, the number of objects with $M_R > -14.5$ after a $V_e$ clipping represents a fair estimate of the number 
of satellites around spirals of $M_R \sim -22$.

\begin{figure}[h]
\begin{raggedright}
\includegraphics[width=1.0\columnwidth]{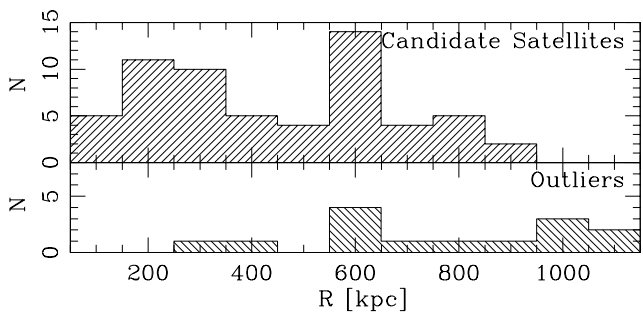}
\par\end{raggedright}

\caption{
  Number of candidate satellites ({\em top} panel) and outliers ({\em bottom} panel)
  as a function of radius.}
\label{fig:densita2}
\end{figure}


\begin{figure}[b]
\begin{raggedright}
\includegraphics[width=1.0\columnwidth]{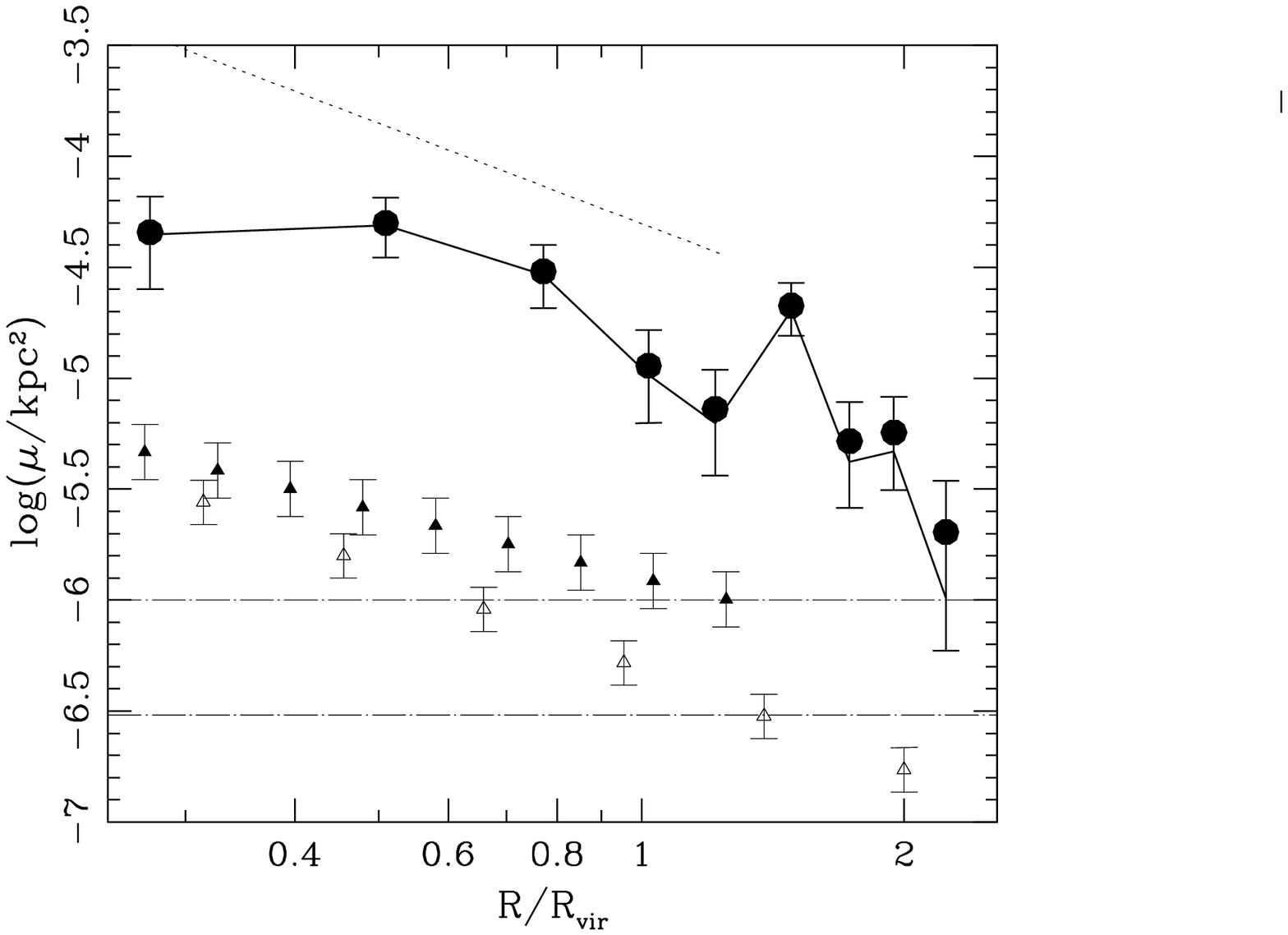}
\par\end{raggedright}

\caption{Surface density of candidate satellites 
({\em filled circles}) as a function of radius. The {\em full} line is the
surface density of satellites after accounting for position
interlopers. The {\em dotted} line indicates the surface density
radial slope of a NFW halo. The {\em full} and {\em open triangles}
represent the surface density of satellites found by Sales et al. (\cite{Sales}) and
Chen et al. (\cite{chen}), respectively. {\em The dashed lines} show the estimate of 
the interlopers velocity for  $\mu = 3 \times
10^{-7}$ kpc$^{-2} $ and $\mu = 10^{-6}$ kpc$^{-2}$.}
\label{fig:densita1}
\end{figure}


\subsection{The radial distribution of satellites}

The radial distribution of satellites was analyzed with bin sizes of
$\sim 100 - 200$~kpc (ten satellites per bin), starting with the total number of satellites 
detected around all primaries. To get a representative number for an individual 
galaxy, the frequencies were later multiplied by the factor of $2/7$, where 
$7$ is the number of primaries in the present work and $2$
is the completeness factor discussed above.
In the full project, we aim to reduce the latter factor down
to $\simeq 1$. In Fig.~\ref{fig:densita1}, we plot the number surface
density $\mu(R)$ before (filled circles) and after (full line) the removal of
interlopers.

The first can be written as 
 
\begin{equation}
\begin{array}{ccc}
\log\ \mu(R)\simeq\log\ \mu_{sat}(R) & = &-4.44+0.18\times(R/100)+\\
 &  & -7.0 \times 10^{-2}(R/100)^{2}\end{array}
 \label{eq:mu},
 \end{equation}
where $R$ is in kpc and $\mu$ in kpc$^{-2}$, the above holding inside $\sim
500$ kpc. We can remove the velocity interlopers, although they
contribute in a negligible way for $R<700$~kpc thus, to a first
approximation, we take Eq.~(\ref{eq:mu}) as the estimate of the
surface density of satellites. 

The projected radial distribution of galactic satellites
around high luminosity primaries has been measured with the SDSS (Chen
et al. \cite{chen}) and the 2dFGRS (Sales et al. \cite{Sales}).
Sales et al. (\cite{Sales}) also distinguished the case of
blue (spiral) and red (elliptical) primaries. In both surveys, the
limiting flux of the satellites is $2$-$3$ magnitudes brighter than
that of our study and halo masses around primaries are a factor
$2$-$5$ smaller than the mass of the halo around our primaries. 
In view of this, the smaller amplitude of the satellite surface density found in the
quoted surveys is unsurprising. It may indicate a bias in the radial distribution of satellites
that depends on their luminosity. We see that $\mu(R)$ is not a power law: there is a central
region of constant value, extending for $\sim 100$~kpc. A similar trend can be seen in the quoted 
surveys, especially for the subsamples of red primaries in 2dFGRS, and high luminosity primaries in SDSS.

Outside $500$~kpc, $\mu(R)$ is subject to the increasingly limited quality of the 
statistics. However, the primary inner kinematics guarantees that the
density profile represents objects related to the primary but not
virialized. The distribution of satellites associated with an $3 \times 10^{12}~
\rm M_\odot$ halo seems to be cut off at about $700$~kpc. The observed
fall-off in number density (see Fig. \ref{fig:densita1}) is real: we
could have detected objects at $R>2 R_{vir}$, if they existed. 

The satellite surface density distribution cannot be reproduced by a projected 
Navarro-Frenk-White (NFW) profile ($\mu \propto R^{-1.5}$) for $0.1<R/R_{vir}<1$
(see Fig.~\ref{fig:densita1}). In other words, it is not compatible
with the predicted density distribution of dark matter haloes in a
$\Lambda$CDM scenario. In particular, the predicted $\sim 200$ subhaloes around a $3\times
10^{12} ~\rm M_{\odot}$ halo exceeds the number of observed satellites
by one order of magnitude, a situation we also find for Local Group
galaxies. However, astrophysical processes in the late stages of galaxy assembly
might depopulate most subhaloes (e.g., Macci\`{o} et al. \cite{maccio}) and the distribution 
of visible satellites might have no resemblance with that of existing DM sub-unities.  

We might try and compare our radial distribution with those predicted by state-of-the-art 
simulations of galaxy formation with models based on cold dark matter (CDM). 
The results of simulations using N-body techniques as well as different semi-analytic models  
of galaxy formation can be found in the literature (e.g., Moore \cite{Moore}; Baugh \cite{Baugh}). 
However, these codes do not produce unique results. We can add as a general comment, that 
these models may have difficulties in explaining the rapid increase in the cumulative number 
of satellites inside $100$~kpc (Macci\`{o} et al.\cite{maccio}).
 
\subsection{Angular distribution of the satellites}

From the $R$-band photometry, we measured the position angle (PA)
of the primaries and derived $\triangle
PA=PA_{satellite}-PA_{galaxy}$ for each satellite. In Fig.~\ref{fig:N&PA}, 
we plot the number of satellites as a function of
$\triangle$PA and in Fig.~\ref{fig:PA&distance} $\triangle$PA as a
function of the distance to the primaries.
We compared the PA distribution to a uniform distribution by means of
a Kolmogorov-Smirnov test. The probability that the PA distribution and the uniform 
distribution have a common origin is  $\sim 98~\%$ (see Fig.~\ref{fig:kstest}).
Hence, the observed distribution is consistent with being isotropic at current 
levels of uncertainty. In any case, it is clear that the distribution of satellites 
around primary galaxy in Figs.~\ref{fig:N&PA} and \ref{fig:PA&distance}
do not show any preferential alignment, i. e. they are distributed
isotropically.
 
\noindent %
\begin{figure}
\begin{centering}
\includegraphics[width=1.0\columnwidth]{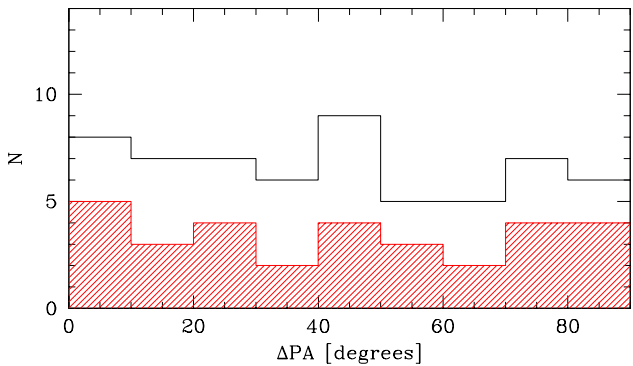}
\par\end{centering}

\centering{}\caption{Number of satellites against  position angle
  differences. The total distribution is given by the upper histogram,
  while that of approaching satellites is represented by the {\em
  unshaded}
  area. In each bin, the rest of the contribution is due to 
 receding satellites ({\em shaded} area). 
\label{fig:N&PA}}
\end{figure}


\begin{figure}
\centering{}\includegraphics[width=1.0\columnwidth]{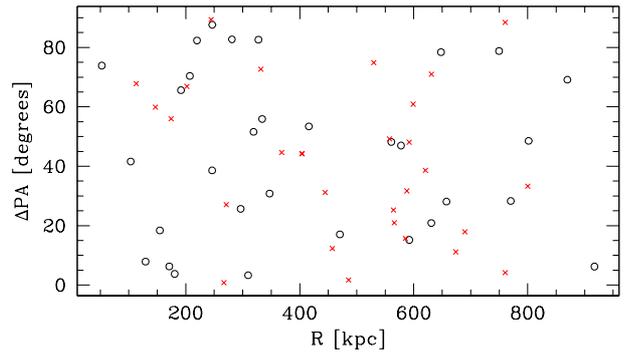}
\caption{Distribution of  satellites around  primary galaxies. Radius
  and position angle with respect to the major axis are shown for
  approaching ({\em open circles}) and receding ({\em crosses}) satellites.
\label{fig:PA&distance}}

\end{figure}

\noindent %
\begin{figure}
\begin{centering}
\includegraphics[width=0.7\columnwidth]{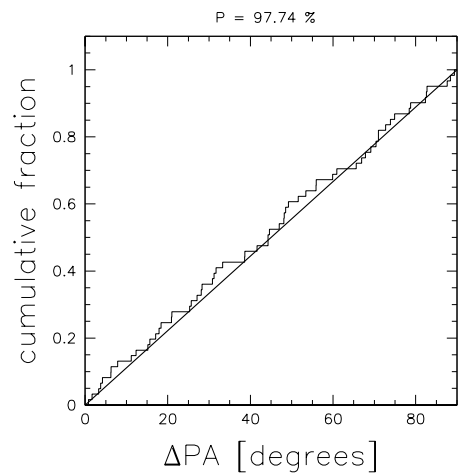}
\par\end{centering}

\centering{}\caption{The cumulate vector of $\Delta \rm PA$ 
({\em broken curve}) is compared to that of a uniform distribution 
({\em straight line}). A Kolmogorov-Smirnov test shows that the probability 
that the two distributions are coming from and identical  underlying one is
$\simeq 98~\%$. \label{fig:kstest}}

\end{figure}

\subsection{Velocity dispersion profile and rotation velocity at large
radii}

\begin{table}[h]
\caption{Velocities and dispersion versus radius. \label{tab:sigma}}

\noindent \begin{raggedright}
\begin{tabular}{rrrrr}
\hline
\hline 
$\left\langle R\right\rangle^{\rm a}$& \multicolumn{2}{c}{$\left\langle V_{\rm rel}\right\rangle^{\rm b}$} & \multicolumn{2}{c}{$\sigma^{\rm c}$}\\
\multicolumn{1}{c}{kpc}& \multicolumn{2}{c}{$~{km~s^{-1}}$} & \multicolumn{2}{c}{$~{km~s^{-1}}$}\\

\hline
142 & 38 &$\pm$ 84 & 267 &$\pm$ 63\\
248 &-59 & 65 & 206  & 49\\
356 &-67 & 51 & 162  & 38\\
521 &101 & 32 & 101  & 24\\
615 &-72 & 99 & 316  & 74\\
799 & 86 & 73 & 232  & 55\\
\hline
\end{tabular}
\par\end{raggedright}
$^{\rm a}$ {\scriptsize average radius of the bin}\\
$^{\rm b}$ {\scriptsize average velocity of the bin and its uncertainty}\\
$^{\rm c}$ {\scriptsize velocity dispersion $\sigma$ of the bin and its uncertainty}\\
\end{table}


\begin{figure}
\begin{raggedright}
\includegraphics[width=1.0\columnwidth]{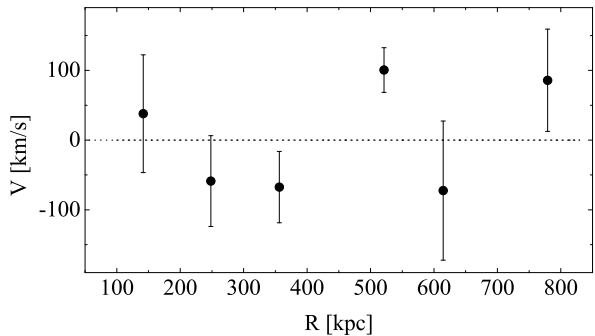}
\par\end{raggedright}

\caption{Mean velocity of satellites as a function of radius.
\label{fig:vmed}}

\end{figure}


\begin{figure}
\begin{raggedright}
\includegraphics[width=1.0\columnwidth]{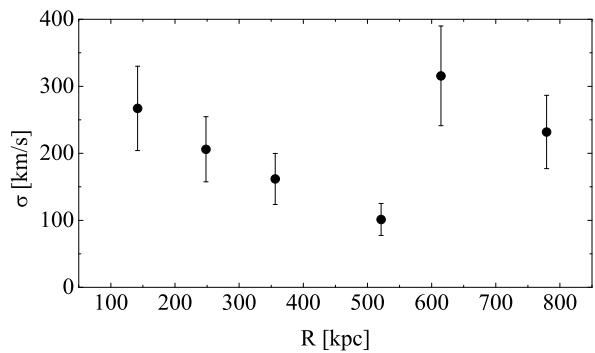}
\par\end{raggedright}

\caption{Velocity dispersion as a function of radius.
\label{fig:smed}}

\end{figure}


We estimate the velocity dispersion $\sigma(R)$ of the seven co-added
systems by binning the satellites into groups of ten objects.
For each bin, we compute the average velocity $\left\langle V_{\rm rel}\right\rangle$ 
and the velocity dispersion $\sigma$.  We report them in Table \ref{tab:sigma}\ and 
plot them in Figs.~\ref{fig:vmed} and \ref{fig:smed}. The average difference 
$\left\langle V_{\rm rel}\right\rangle$ 
is compatible with zero at any radius (see Fig.~\ref{fig:vmed}). In addition, along 
the line-of-sight the velocities are isotropic. This is confirmed by the finding that 
$\left\langle V_{\rm rel}\right\rangle \sim 0.2 \sigma$ at any radius
(see Table~\ref{tab:sigma}). The flat radial trend also implies  that the satellites 
system has a negligible rotation.

In view of the considerably large r.m.s. of $\sigma(R)$,  we chose  the simplest linear function  
to fit data  from 100 kpc to 500 kpc

\begin{equation}
\sigma(R)= [265 - 0.35 (R - 100)] ~{\rm km~s^{-1}},
\end{equation}
where $R$ is in kpc. 

The inner RCs of the primaries indicate, following Salucci et al. (\cite{URCII}), for our co-added system a
virial mass of $\sim 3.5 \times 10^{12}~\rm M_{\odot}$ and a virial radius of $400$ kpc. Inside this 
radius, $\sigma$ measures the kinetic energy of a unit mass test particle in gravitational equilibrium
inside the galaxy potential.  At larger radii in the range $400< R< 800$,~kpc we have two cases: 
1) $\sigma$ still measures the kinetic energy, but of objects out of equilibrium or 2) $\sigma$ 
is totally contaminated by velocity interlopers.

The velocity dispersion measures the circular velocities well beyond the extent of the rotation curves.
We assume that the satellites have isotropic motions. This assumption is supported (albeit not proven) 
by the isotropic angular distribution and velocity dispersion along the line of sight. 
The co-addition procedure would of course also smear out any velocity
anisotropies if they had different structure galaxy by galaxy. 
In this framework, we derive the circular velocity $V= R d \Phi/dR $
where $\Phi$ is the gravitational potential, from the velocity dispersion
in two different ways.

First, we apply the usual argument that velocity dispersion has three
degrees of freedom while rotation has just two, and we set $V(R)=
(3/2)^{1/2} \sigma(R)$. Rotation velocities computed this way are plotted 
in Fig.~\ref{fig:bin-sat1} together with the inner RCs.

Second, we adopt the fully correct procedure: we solve the Jeans
equation for the 3D distribution of satellite $n_{\rm s}(R)$ that we
get from an Abel-integral deconvolution of $\mu(R)$. We note that in
this pilot study the quality of the statistics is still low and this 
is done just to illustrate the feasibility of the full project 
(which involves several times more data).  The deconvolution yields
\begin{equation}V(R)= k(R) \sigma(R) \end{equation}
with $k=(-dlog \ n_s(R) /dlog\ R - 2\ dlog\ \sigma/dlog\ R)^{0.5}$. 
For $r<400$~kpc, i.e. for velocity dispersions
of the three innermost bins we find  $k=1.0, 1.4, 1.8$. For the
outer three bins in which very likely the Jeans equation cannot be applied,
we allow $V$ to range between $(3/2)^{1/2} \sigma $ and $\sigma$ (an object radially 
infalling) and we do not estimate a value.
Fig.~\ref{fig:bin-sat2} shows the inner RCs together with velocities
computed with this second procedure. 

The last two bins illustrate a steep, large increase in the velocity dispersion. This 
is likely to  indicate that most, perhaps all objects, are not physically bound to the primary 
galaxy. The detected sharp decrease in the number of true  satellites  might indicate  at ~$600~\rm kpc$ 
the edge of the dark matter halo. With the full sample that is four times larger, we hope of course 
to have consequently a more accurate determination of the mass distribution.

\subsection{Mass modeling}

The radial profiles of circular velocities shown in Figs.~\ref{fig:bin-sat1}
and \ref{fig:bin-sat2} were fit with the universal rotation curve (URC)
with $\left\langle R_\mathrm{D}\right\rangle = 6.2$~kpc and halo mass
$M_{vir}=3.5 \times 10^{12}~\rm M_{\odot}$ that closely fits the inner RCs.
The URC has no free parameters to fit the present outer data, therefore its success 
is remarkable. The URC depends only on the disk scale-length
$R_\mathrm{D}$ and the halo mass $M_{\rm vir}$, which also controls the disk mass
(Shankar et al. \cite{Shankar}). The URC mass model fits well the data points in both cases 
for $M_{vir}=3.5 \times 10^{12}~\rm M_{\odot}$.

\begin{figure}[h]

\includegraphics[width=1.0\columnwidth]{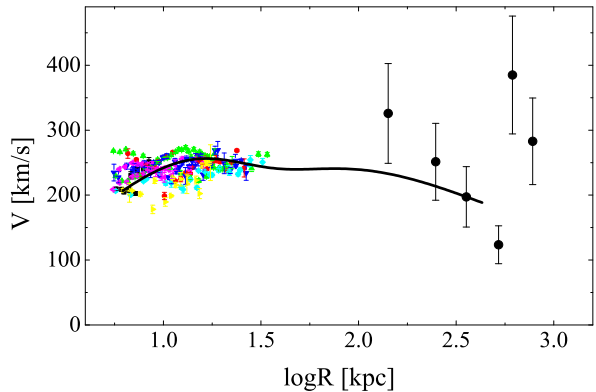}

\caption{RCs of primaries and circular velocities  obtained from  the velocity dispersions of the satellites
({\em black squares}) ($V(R)= (3/2)^{1/2} \sigma(R)$, see text for details). 
The URC is shown (as a {\em black line}) out to the virial radius.
\label{fig:bin-sat1}}

\end{figure}

\begin{figure}[h]
\begin{raggedright}
\includegraphics[width=1.0\columnwidth]{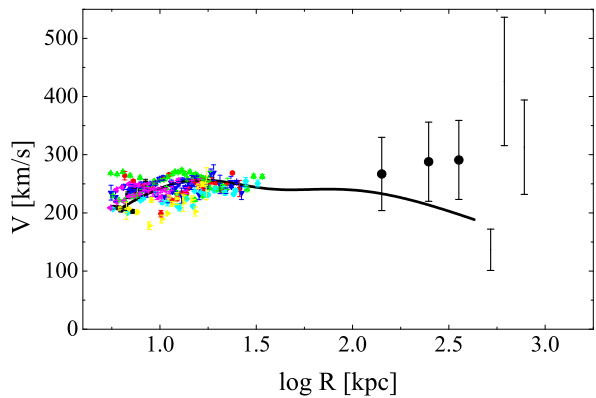}
\par\end{raggedright}

\caption{RCs of primaries and circular velocities  obtained from velocity dispersions  and jeans Equation 
({\em black squares}, see text for details). The URC is shown (as a {\em black line}) out to the virial radius.
\label{fig:bin-sat2}}

\end{figure}

\section{Discussion and conclusions}

In this pilot project, we have searched for satellites around seven spiral galaxies 
of comparable luminosity ($M_R \sim -22$). We have found 61 objects that, based 
on their velocity relative to the primary galaxies, are their satellites. By combining 
results, we have investigated several properties of these satellites, which allow us to draw 
a number of conclusions.

We find that our target dark matter haloes have about 15 satellites (down to
$M_{R}=-16$) inside their virial radius.  Their measured inner 
kinematics show that, as intended, they have similar halo masses, of about 
$3.5 \times 10^{12}~\rm M_{\odot}$
and the embedded luminous spirals have
luminosities $M_{R} \simeq -22$. Therefore this result is comparable
to what we observe for M31 and the Milky Way (MW) in the Local Group. 
As in those cases, this small number of satellites represents a problem
for ``naive'' $\Lambda$CDM theories, which predict an order of magnitude
larger number of objects orbiting around dark haloes of this size. To
reconcile observations and theory, one has to assume that most of the
galaxy subhaloes have become unobservable. This possibility, related also to the apparent 
lack of subhaloes around the Milky Way, is being investigated (e.g. Kravtsov et al. \cite{Kravtsov}; 
Libeskind et al. \cite{Libeskind}).
Could a long list of physical processes occurring in subhaloes  during their history  
jeopardize (while perhaps also  introducing a radial bias) their ability to host luminous objects? 
In this respect, we note that our primaries  have a mass 3-4 times higher that of the MW and 
the combined sample  has already $10^2$ objects and in the future will have ten times more.  
We will be able to investigate the "missing subhaloes problem" also in subhaloes with a mass  
1 order of magnitude  larger than  that of the Large Magellanic Cloud (LMC), and therefore 
in subhaloes that should host normal and not dwarf galaxies. It will be interesting to see if, 
at these  high masses, the problem remains, and can be tackled by the explanations put forward 
in the case of MW.

By considering a composite system including all detected satellites,
we are able to study their radial distribution with respect to the 
central galaxy. The surface density has a shallower profile than 
found in previous surveys, based on fewer satellites per primary.
We propose that this is due to a bias in the radial distribution 
of satellites that depends on their luminosity.

We also find that the angular distribution of satellites around
primary spirals is isotropic. Furthermore, 
the relative velocity along the line of sight is isotropically
distributed around zero. Based on this observation, we 
can use the velocity dispersion profile to estimate rotation 
velocities far beyond the extent of the inner rotation curves.

An additional result of this study is that the RCs of the primaries appear 
very similar. Because we targeted objects with the same absolute luminosity, 
our finding confirms that RCs mainly depend on an object's mass. It follows 
that the mass distribution inside the optical region is the same for all
our objects. It also justifies combining the satellite relative motions 
to derive the velocity dispersion and probe the mass distribution of 
a ``stacked'' halo. The outer RC obtained from the velocity dispersion 
results is smooth and continuous with respect to the inner measured RCs and, 
over two orders of magnitude in distance, is very well reproduced by the S07 universal
mass model that holds for spirals. In this paper, the model is tested for the first 
time with kinematical measurements out to the virial radius of the galaxy.
The technique we have applied is therefore very promising. 

In the future, we plan to test our findings by enlarging our sample 
of primary galaxies. We estimate that by co-adding the satellite distribution 
and kinematics of $\sim 20$ systems of a same high mass we can infer the
mass distribution of the underlying DM halo. The full project aims to replicate 
this procedure for haloes of intermediate and low mass ($8 \times 10^{11}~\rm M_{\odot}$ 
and $2\times 10^{11}~\rm M_{\odot}$ respectively), to derive the mass
dependence of the DM distribution.

\begin{acknowledgements} 
We thank Teresa Brainerd for kindly providing the catalog of the primaries with satellites obtained
from the SDSS survey.We thank Antonaldo Diaferio for kindly computing the caustics curves to
identify satellites and outliers. We are grateful to the anonymous referee for valuable comments 
that truly improved the paper. AP acknowledge findings through grant  CPDA089220/08 by Padua University. 
We are pleased to thank Michael West for improving the style of the manuscript and for the helpful comments. 
We thank Ivo Saviane for his reading of the manuscript that considerably improved the presentation of this work.  
The Sloan Digital Sky Survey (SDSS) is a joint project of The University of Chicago, Fermilab, the Institute 
for Advanced Study, the Japan Participation Group, The Johns Hopkins University, Los Alamos National Laboratory, 
the Max-Planck Institute for Astronomy (MPIA), the Max-Planck Institute for Astrophysics (MPA), New Mexico State 
University, Princeton University, the United States Naval Observatory, and the University of Washington. Apache
Point Observatory, site of the SDSS telescopes, is operated by the Astrophysical Research Consortium (ARC). 
Funding for the project has been provided by the Alfred P. Sloan Foundation, the SDSS member institutions, 
the National Aeronautics and Space Administration, the National Science Foundation, the US Department of Energy, 
the Japanese Monbukagakusho, and the Max Planck Society. The SDSS Web site is http://www.sdss.org/.

\end{acknowledgements}

\end{document}